\documentclass[review]{elsarticle}
\graphicspath{ {./figures/} }
\usepackage{hyperref}
\usepackage{float}
\usepackage{verbatim} %comments
\usepackage{apalike}
\restylefloat{figure}
\restylefloat{table}

\usepackage[linesnumbered,ruled,vlined]{algorithm2e}

\journal{Engineering Applications of Artificial Intelligence}

\biboptions{authoryear}

\begin{document}
\begin{frontmatter}

\begin{titlepage}
\begin{center}
\vspace*{1cm}

\textbf{ \large NeSHFS: Neighborhood Search with Heuristic-based Feature Selection for Click-Through Rate Prediction}

\vspace{1.5cm}

% Author names and affiliations
Dogukan Aksu$^a$ (d.aksu@istanbul.edu.tr), Ismail Hakki Toroslu$^b$ (toroslu@ceng.metu.edu.tr), Hasan Davulcu$^c$ (hdavulcu@asu.edu) \\

\hspace{10pt}

\begin{flushleft}
\small  
$^a$ Department of Computer Engineering, Istanbul Unıversity-Cerrahpasa, Istanbul, Turkey \\
$^b$ Department of Computer Engineering, Middle East Technical University, Ankara, Turkey \\
$^c$ Computer Science and Engineering, Arizona State University, Tempe, AZ, USA 

\begin{comment}
Clearly indicate who will handle correspondence at all stages of refereeing and publication, also post-publication. Ensure that phone numbers (with country and area code) are provided in addition to the e-mail address and the complete postal address. Contact details must be kept up to date by the corresponding author.
\end{comment}

\vspace{1cm}
\textbf{Corresponding Author:} \\
Ismail H. Toroslu \\
Dept. of Computer Eng. METU, Ankara, Turkiye \\
Email: toroslu@ceng.metu.edu.tr

\end{flushleft}        
\end{center}
\end{titlepage}

\title{NeSHFS: Neighborhood Search with Heuristic-based Feature Selection for Click-Through Rate Prediction}

\author[label1]{Dogukan Aksu}
\ead{d.aksu@istanbul.edu.tr}
%\ead{dogukanresearch@gmail.com}

\author[label2]{Ismail Hakki Toroslu \corref{cor1}}
\ead{toroslu@ceng.metu.edu.tr}

\author[label3]{Hasan Davulcu}
\ead{hdavulcu@asu.edu}

\cortext[cor1]{Corresponding author.}
\address[label1]{Department of Computer Engineering, Istanbul University-Cerrahpasa, Istanbul, Turkey}
\address[label2]{Department of Computer Engineering, Middle East Technical University, Ankara, Turkey}
\address[label3]{Computer Science and Engineering, Arizona State University, Tempe, AZ, USA}

\begin{abstract}
Click-through-rate (CTR) prediction plays an important role in online advertising and ad recommender systems. In the past decade, maximizing CTR has been the main focus of model development and solution creation. Therefore, researchers and practitioners have proposed various models and solutions to enhance the effectiveness of CTR prediction. Most of the existing literature focuses on capturing either implicit or explicit feature interactions. Although implicit interactions are successfully captured in some studies, explicit interactions present a challenge for achieving high CTR by extracting both low-order and high-order feature interactions. Unnecessary and irrelevant features may cause high computational time and low prediction performance. Furthermore, certain features may perform well with specific predictive models while underperforming with others. Also, feature distribution may fluctuate due to traffic variations. Most importantly, in live production environments, resources are limited, and the time for inference is just as crucial as training time.  Because of all these reasons, feature selection is one of the most important factors in enhancing CTR prediction model performance. Simple filter-based feature selection algorithms do not perform well and they are not sufficient. An effective and efficient feature selection algorithm is needed to consistently filter the most useful features during live CTR prediction process. In this paper, we propose a heuristic algorithm named Neighborhood Search with Heuristic-based Feature Selection (NeSHFS) to enhance CTR prediction performance while reducing dimensionality and training time costs. We conduct comprehensive experiments on three public datasets to validate the efficiency and effectiveness of our proposed solution.
\end{abstract}


\end{frontmatter}

\section{Introduction}
\label{introduction}

The development of the Internet has provided various platforms for reaching potential customers, including search engines, e-commerce websites, mobile apps, and social media platforms. Consequently, online advertising has emerged as a dominant sector in advertising \citep{yang2022click}.

Predicting the probability of a user clicking on a specific ad is critical in online advertising. There are various performance metrics such as Click-Through-Rate (CTR) and Convergence-Rate (CVR) used to determine the relevance between ads and users.

A lot of practitioners and researchers have explored various approaches and techniques to enhance Click-Through Rate (CTR) prediction over the past decades. Many of these successful methodologies leverage machine learning, utilizing historical click data to forecast future user behavior \citep{yang2024new, duan2024attacking, demsyn2024position, ma2024deep, wang2024multi}. Consequently, the selection and quality of features play a pivotal role in determining the performance of these prediction models. Features serve as the building blocks upon which these models operate, influencing their ability to accurately capture patterns and relationships within the data. Therefore, understanding the relevance and impact of different features is crucial for optimizing the predictive power of CTR models.

Novel Click-Through Rate (CTR) prediction techniques typically incorporate two key components: the embedding layer and the feature interaction layer.

The embedding layer transforms categorical features into dense, low-dimensional representations, known as embeddings. These embeddings capture the underlying relationships and similarities between categorical values, enabling the model to effectively learn from categorical data.

On the other hand, the feature interaction layer facilitates the modeling of interactions between different features, both categorical and numerical. This layer allows the model to capture complex relationships and dependencies between features, which are crucial for the accurate prediction of click-through rates.

Overall, the success of CTR prediction models is commonly evaluated using the area under the curve (AUC) metric, as it is a binary-class classification problem. Even marginal improvements in AUC, such as a 0.1 \% increase, are considered significant, given the massive scale of online advertising platforms. Typically, a small improvement in AUC can translate to substantial gains in revenue, potentially amounting to hundreds of thousands of dollars, considering the vast number of users and clicks \citep{song2024multi, wu2024metasplit, zhang2024incmsr}. 

In real-world applications, a diverse array of models is employed to predict CTR and CVR  within production environments. Given the vast scale of users and items involved, it's estimated that nearly a billion exposures occur daily in live environments \citep{ji2021reinforcement, dong2024tales}. These models are tasked with swiftly handling the massive influx of traffic, typically within ten milliseconds (ms), considering the constant stream of user interactions captured at this frequency \citep{de2024drivers}.

In live environments where resources are constrained, making rapid decisions is just as crucial as accurate CTR predictions \citep{veneranta2024optimization, guo2019pal}. Failure to respond within ten ms may result in decisions being made by rule-based models, which often lag behind deep learning models in performance \citep{mukherjee2017apriori}. Consequently, the effectiveness of recommending relevant ads to users diminishes significantly, leading to potential revenue losses.

Furthermore, both the inference time and training time of models directly impact CTR outcomes in live environments. Lengthy training periods limit resources available for experimentation in offline production. Additionally, the continued use of the most recently trained model may hinder adaptability to changes in traffic patterns. Thus, minimizing both training and inference times is critical for achieving optimal performance in real-world scenarios.

As a result, the feature selection is very important in this domain. Furthermore, many researches reveal that feature selection not only decreases the training and inference time but also improves the model's success in predicting CTR \citep{li2017feature}. Since feature sets can often be extensive, identifying the optimal subset of features essentially entails exploring every possible combination. However, due to the exponential time complexity of this task, practical heuristic-based approaches are necessary to identify subsets that yield results approaching the optimal solution.

In this paper, we propose NeSHFS, a heuristic-based feature subset selection technique, based on ranking, removing features recursively, and searching for better solutions around the neighborhood to obtain a high prediction capability of CTR models. 
Our idea has been inspired from the {\it grid search} hyperparameter optimization technique \citep{bergstra2012random}. The goal of grid search is to find the combination of hyperparameters that yields the best performance according to a specified evaluation criterion. We aim to apply grid search like an idea in feature selection, extended with local neighbourhood search. 

Instead of utilizing generic heuristics, we have developed a customized heuristic that fits to this problem. Our heuristic method both evenly explores whole search space and then exploits the areas with more promising results. We utilize the DeepCTR framework, and an 80$/$10$/$10 split for training, validating, and testing to validate the efficiency of the proposed solution and predict click-through-rate in the experiments. 

The main contributions of this paper can be summarized as follows:
\begin{itemize}
    \item We introduce a novel feature selection approach, named NeSHFS, aimed at efficiently predicting CTR by rapidly selecting a near optimal subset of features.
    \item Due to reducing the set of features, NeSHFS decreases deep CTR model training time significantly, so that the model can be retrained continuously.  Moreover, it also reduces the inference time to get more traffic.
    \item NeSHFS is suitable to be utilized by any deep CTR model based due to its modular structure.
    \item We have also verified the efficiency of the proposed solution on a comprehensive testbed with three public datasets, namely: Huawei Digix 2022, Criteo, and Avazu. 
\end{itemize}

The rest of the paper is organized as follows: Section \ref{related_works} provides related works based on feature selection techniques and CTR prediction models. Section \ref{proposed_methodology} presents the proposed methodology. Section \ref{evaluations_and_results} demonstrates the outcomes of the conducted experiments. Section \ref{discussion} contains discussion. Section \ref{conclusions} includes conclusions.

\section{Related Works}
\label{related_works}

Feature selection is a crucial technique for identifying a representative subset of original features in data, thus reducing computational time while preserving or enhancing prediction results. In general, feature selection techniques fall into three categories: filter, wrapper, and embedded \citep{chandrashekar2014survey}. However, finding the best feature subset is an NP-hard problem due to the exponential search space including all possible subsets \citep{aksu2022mga}. Thus, heuristic approaches are preferred over optimal ones, especially in scenarios with a large number of features, due to their lower computational cost \citep{theng2024feature}.

Various feature selection methods have been proposed in the literature. For instance, Aksu and Aydin \citep{aksu2022mga} introduce the Modified Genetic Algorithm (MGA) for selecting near optimal set of features for intrusion detection problem. Liu et al. \citep{liu2024feature} develop Feature Subset Selection using Mutual Information (FSM) to identify optimal feature subsets. Guyon and Elisseeff \citep{guyon2003introduction} advocate for the use of linear classifiers, such as Support Vector Machines (SVMs), with ranking-based or nested subset selection methods. Additionally, Ma et al. \citep{ma2024class} propose a novel class-specific feature selection algorithm using fuzzy information-theoretic metrics, which tailors feature subsets for each class to enhance classification performance.

Li et al. \citep{li2017feature} discuss feature selection techniques from a data perspective, while Thakkar and Lohiya \citep{thakkar2023fusion} employ a fusion of statistical importance for feature selection in intrusion detection problem.

Moreover, feature selection techniques have been used in many other domains such as stock market price prediction (\citep{khan2020arima}). For example, Aloraini (\citep{aloraini2015penalized}) employs ensemble feature selection based on Pearson and Spearman correlations, demonstrating better prediction results compared to single feature selection techniques.

CTR prediction is a critical task in online advertising, and various deep learning models have been proposed to enhance prediction performance. Guo et al. (\citep{guo2017deepfm}) propose DeepFM for maximizing CTR in recommender systems, while Liao et al. (\citep{liao2022maskfusion}) present MaskFusion to improve deep CTR models' performance. Additionally, Tian et al. (\citep{tian2023eulernet}), Wang et al. (\citep{wang2020click}), Zhang et al. (\citep{zhang2023fibinet++}), Chen et al. (\citep{chen2019flen}), Song et al. (\citep{song2019autoint}), and Deng et al. (\citep{deng2023contentctr}) propose different deep learning models for CTR prediction, each with its unique approach.

Despite the widespread application of feature selection techniques, there is a lack of studies focusing on improving the CTR prediction performance of deep models through feature selection. Therefore, in this paper, we address this gap by proposing a novel feature selection technique, NesHFS, aimed at enhancing the CTR prediction capabilities of models.

\section{Proposed Methodology}
\label{proposed_methodology}
Identifying the optimal subset of features for achieving the highest AUC score requires evaluating all feature subsets, hence, it takes an exponential amount of time. Therefore, rather than looking for the global optimum solution, it is more practical to efficiently determine a local optimum feature set that generates an AUC score close to the best one. Our proposed method is a customized heuristic to explore a very small portion of all possible feature subsets in order to obtain strong local optima. This exploration is done by utilizing the significance scores of features and by defining neighborhood concepts among feature subsets based on the scores of features. 

In the NeSHFS method, we first calculate feature scores using chi-square and ANOVA methods for sparse and dense features, respectively. Subsequently, these sparse and dense features are ranked based on their scores independently. Next, a certain number of features are removed from both feature sets, with the specific numbers determined according to the sizes of the sparse and dense feature sets. Following each removal, CTR prediction is conducted on a dataset containing the remaining features, utilizing models such as DeepFM and FiBiNet. This process iterates until no features remain.

Following this global search, a refined search is conducted to explore potentially better feature sets. This involves selecting the top k (in our case, 3) feature sets that produced the highest AUC scores during the global search. After that, small neighborhoods of these sets are generated by iteratively adding or removing features one by one. These refined feature sets are then subjected to the same CTR prediction process. As a result, the feature set producing the highest AUC score from the final neighborhood search step is returned, which corresponds to a local optimum.

\begin{figure}[H]
  \centering
  \includegraphics[width=\linewidth]{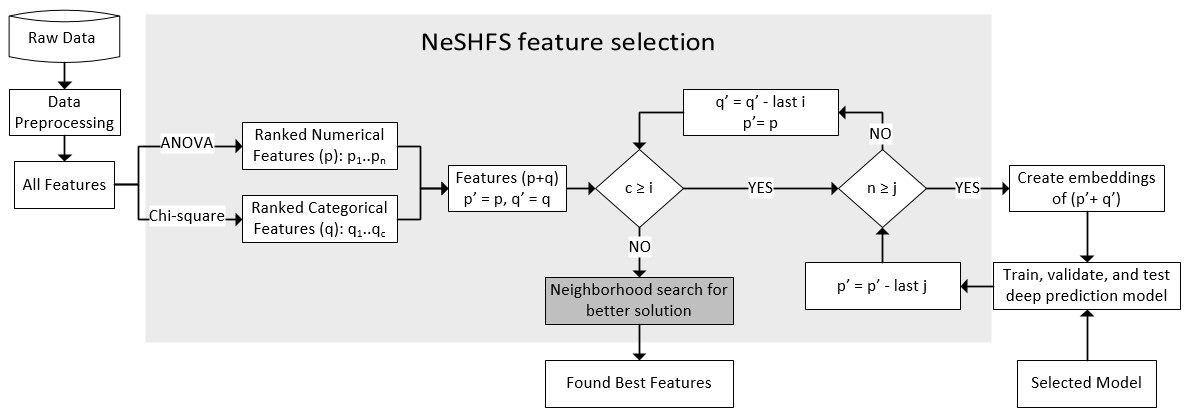}
  \caption{Flowchart of the proposed solution.}
  \label{flowchart}
\end{figure}

The architecture of the proposed NeSHFS is given in Fig. \ref{flowchart}.

The  main components of NeSHFS are as follows:   
\begin{enumerate}
    \item Preprocessing Module: In this module, first, raw data is cleaned and useful features are selected. Then, ANOVA and chi-square scores of numerical and categorical features are calculated, and they are separately ranked according to their scores.
%    \item Feature scoring \& ranking: Features are ranked by their ANOVA and chi-square scores for dense and sparse features, respectively.
    \item Model Training and CTR Prediction Module: This module is used to determine the success of the feature subset generated by NeSHFS. One of the well-known models, such as DeepFM and FiBiNet, are used with the dataset of the selected subset of features. Then, using the test data set, the AUC scores corresponding to the feature subset is calculated.
    \item Neighborhood search: In order to potentially find a better feature subset, its neighbours are also generated and processed. Neighbours of selected feature subset are determined by up and down neighborhood search techniques, which are described in detail below. This process involves adding or deleting one feature at a time. In our experiments, we have searched 3 neighbours in each direction. For each neighbour subset, its success is determined using the same model training and CTR prediction method.
    \item Global Feature Selection Module: This is the main part of the whole process. It is aimed to search a wide area of search space while going into detailed search for promising parts. This is done by utilizing scores of features. Initially, all features are arranged in descending order of scores, with separate lists for numerical and categorical features. The module operates through nested iterations: an outer loop for removing categorical features and an inner loop for removing numerical features. In the first iteration of the outer loop, the entire feature set is utilized for CTR prediction. Subsequently, in the inner loop, a selected number of features with the lowest scores are systematically deleted from the numerical features, and the resulting subset is used for CTR prediction. This process continues until there are not sufficient numerical features left. At this point the inner loop ends. Similarly, in each iteration of the outer loop, a selected number of categorical features with the lowest scores are removed. Furthermore, each iteration of the inner loop begins with the complete numerical feature set. The outer loop terminates when there are not enough features remaining.
\end{enumerate}

NeSHSF has a parametric structure so, it can be adapted to different data sets. The definition of the parameters utilized in our experiments is shown in Table \ref{parameter_table}. 

    \begin{table}[H]%\scriptsize   %\footnotesize \tiny \small
        \caption{Parameter definitions.}
        \label{parameter_table}
        \centering
        \begin{tabular}{ll}
        \hline
        \textbf{Parameter}  &  {\textbf{Definition}}     \\
        \hline
        \textbf{$p$}      & numerical features from $p_1$ to $p_n$.  \\ 
        \textbf{$q$}      & categorical features from $q_1$ to $q_c$.  \\
        \textbf{$i$}      & number of features will be removed from categorical features. \\ 
        \textbf{$j$}      & number of features will be removed from numerical features. \\
        \textbf{$u$}      & size of up search.    \\  
        \textbf{$d$}      & size of down search.    \\ 
        \hline
       \end{tabular} 
    \end{table}

 As shown in Table \ref{parameter_table}, numerical features are denoted by $p$ and categorical features are represented by $q$. The parameters $i$ and $j$ indicate the number of features to be removed in iterations for categorical and numerical features, respectively. 
 
 In the context of the neighborhood search, $u$ signifies the size of the up search, while $d$ corresponds to the size of the down search.
 
 The pseudo-code of the neighborhood search is given in the Algorithm \ref{algo1}.
 
\begin{center}
    \begin{algorithm}[H]%\tiny   %\footnotesize \tiny \small
        \label{algo1}
            \SetKwInOut{Input}{Require}
            \SetKwInOut{Output}{Ensure}
            \Input{$p$, $q$, $i$, $j$, $k$, $u$, $d$}
            \Output{Found Best Feature Subset: $X_{best}$.}
            Rank results by AUC. \\
            $k$ $\gets$ top $k$ feature sets.\\
            \While{$k$ is not empty}{ % top n feature set için
                $selected_{set}$ $\gets$ Pop feature set from k. \\
                $p$ $\gets$ categorical features from the $selected_{set}$. \\
                $q$ $\gets$ numerical features from the $selected_{set}$. \\
                $X_{ubest}$ $\gets$ up-search($p$, $q$, $u$). \\
                $X_{dbest}$ $\gets$ down-search($p$, $q$, $d$). \\
            }
            return $X_{best}$.\\
        \caption{General search.}
    \end{algorithm}
\end{center}

Initially, a dictionary data structure is created, where feature subsets serve as keys and their corresponding AUC scores as values. Subsequently, the dictionary is sorted based on AUC scores, and the top $k$ results are extracted and stored in a list $k$ containing feature subsets.

Until $k$ is empty, the following steps are executed: a feature set is selected from $k$ and assigned to $selected_{set}$. Categorical and numerical features are then identified, and up and down searches are applied sequentially.

Finally, the feature subset with the highest AUC value is returned. Pseudo-codes for the up-search and down-search algorithms are provided in Algorithm \ref{algo2} and Algorithm \ref{algo3}, respectively.

\begin{center}
    \begin{algorithm}[H] %\tiny %\small %\tiny   %\footnotesize 
        \label{algo2}
            \SetKwInOut{Input}{Require}
            \SetKwInOut{Output}{Ensure}
            \Input{$p$, $q$, $u$}
            \Output{Best Feature Subset: $X_{ubest}$}
            \While{u > 0}{
                Removed numerical features: $r_p$ \\
                Removed categorical features: $r_q$ \\
                \If{$|r_p|$ is equal to $0$}{
                    feat $\gets$ pop feature from $r_q$  \\
                    add feat to $q$.  \\
                    pop $j$-features from $p$, and add them to $r_p$. \\
                }
                \Else{
                    pop feat from $r_p$, and add feat to $p$. \\
                }
                e $\gets$ Create feature embeddings (p + q). \\
                model $\gets$ deep prediction model. \\
                Train, validate, and test model with e. \\
                u $=$ u - 1. \\ 
                }
            return $X_{ubest}$ \\
        \caption{Up-neighborhood search.}
    \end{algorithm}
\end{center}

Algorithm \ref{algo2} outlines the following steps, executed until the iteration number is reached:
\begin{enumerate}
    \item Identification of removed numerical and categorical features.
    \item If there are no removed numerical features, a recently removed feature from categorical features is added back to categorical features, while $j$ features are removed from numerical features and added to the removed numerical features. Otherwise, a feature is removed from removed numerical features and added to removed numerical features.
    \item Creation of embeddings for the updated numerical and categorical features.
    \item Training, validation, and testing of a selected deep prediction model using the created embeddings.
    \item Return of the feature subset with the best AUC score.
\end{enumerate}

\begin{center}
    \begin{algorithm}[ht] %\tiny   %\footnotesize \tiny \small
        \label{algo3}
            \SetKwInOut{Input}{Require}
            \SetKwInOut{Output}{Ensure}
            \Input{$p$, $q$, $d$}
            \Output{Best Feature Subset $X_{dbest}$}
            \While{$d$ > 0}{
                Removed numerical features: $r_p$ \\
                Removed categorical features: $r_q$ \\
                feat $\gets$ pop feature from $p$. \\
                Add feat to $r_p$. \\
                \If{$|p|$ is equal to $0$}{
                    feat $\gets$ pop feature from $q$  \\
                    add feat, and all $r_p$ to $r_q$, and $p$, respectively.  \\
                }
                $e$ $\gets$ Create feature embeddings ($p$ + $q$). \\
                model $\gets$ deep prediction model. \\
                Train, validate, and test the model with $e$. \\
                $d$ $=$ $d$ - 1. \\ 
                }
            return found best feature subset. \\
        \caption{Down-neighborhood search.}
    \end{algorithm}
\end{center}

Algorithm \ref{algo3} continues to search for the best feature subset until the iteration number is reached. The steps involved are as follows:
\begin{enumerate}
    \item Identification of removed numerical and categorical features.
    \item Removal of a feature from numerical features, which is then added to the removed numerical features. If all numerical features are removed, a feature from categorical features is removed, and all removed numerical features are added back to numerical features.
    \item Generation of embeddings for the updated numerical and categorical features.
    \item Training, validation, and testing of a selected deep prediction model using the generated embeddings.
    \item Return of the feature subset with the best AUC score.
\end{enumerate}

\section{Evaluations and results}
\label{evaluations_and_results}
This section presents the outcomes of the proposed NeSHFS method, focusing on Click-Through Rate (CTR) prediction. Utilizing the DeepCTR framework, we evaluate the efficiency of our proposed method using benchmark datasets, including the Huawei Digix 2022 dataset, the Criteo dataset, and the Avazu dataset. Our primary evaluation metric employs the DeepFM model for CTR prediction. Additionally, we employ the FiBiNeT model to further validate our proposed method as another deep predictive model, specifically on the Huawei Digix 2022 dataset. Furthermore, to validate the efficiency of our feature selection technique, we compare the proposed solution with the Genetic Algorithm (GA). 

All experiments are conducted on an NVIDIA GTX 950M GPU. A batch size of 256 is utilized, and approximately 1 million randomly selected unbalanced data points are employed due to hardware limitations. To evaluate the efficiency of the proposed method and predict Click-Through Rates (CTR), an 80/10/10 split is used for training, validation, and testing, respectively.

To prevent over-fitting, we use early stop, and its patience is set to three on validation loss. Other parameters, such as the number of hidden units in the Deep Neural Network (DNN), l1 regularization, l2 regularization, and dropout, were used with their default values.

Each experiment was conducted five times, and the average results are presented in the result tables alongside the corresponding number of features. The numbers in parentheses within the number of features column represent the number of numerical and categorical features, respectively.

During the experiments, $i$ and $j$ values are set to 5 and 3, respectively. Furthermore, defined parameters and their values are summarized in Table \ref{params}. The values of $p$ and $q$ vary depending on the selected dataset. Parameters $u$, $d$, and $k$ are arbitrarily chosen as three. 

    \begin{table}[H]%\scriptsize   %\footnotesize \tiny \small
        \caption{Values of parameters.}
        \label{params}
        \centering
        \begin{tabular}{llllllll}
        \hline
        \textbf{Dataset} & \textbf{$p$} & \textbf{$q$} & \textbf{$i$} & \textbf{$j$} & \textbf{$u$} & \textbf{$d$} \\ %% & \textbf{$k$} \\
        \hline
        \textbf{Digix}  & 3  & 22 & 5 & 3 & 3 & 3 \\ %%& 3 \\
        \textbf{Criteo} & 13 & 26 & 5 & 3 & 3 & 3 \\ %%& 3 \\
        \textbf{Avazu}  & 0  & 24 & 5 & 0 & 3 & 3 \\ %%& 3 \\
        \hline
       \end{tabular} 
    \end{table}

Table \ref{t7} demonstrates the order of features based on their importance score. The most important categorical features are slot\_id, C12, and device\_id on the Digix dataset, Criteo dataset, and Avazu dataset, respectively. The most influential numerical features are u\_refreshTimes and I3.

\begin{table}[H]\scriptsize   %\footnotesize \tiny \small
    \caption{ Features of Digix, Criteo, and Avazu dataset.}
    \label{t7}
    \centering
       \begin{tabular}{llll}
        \hline
        \hline
        \multicolumn{4}{l}{(a). Categorical features. }\\
        \hline
        \#Importance  & Digix Features     & Criteo Features          & Avazu Features    \\
        1     & slot\_id                  & C12                       & device\_id    \\
        2     &  adv\_prim\_id            & C16                       & C14   \\    
        3     &  adv\_id                  & C4                        & app\_id    \\  
        4     &  user\_id                 & C10                       & device\_ip    \\   
        5     &  pt\_d                    & C21                       & C17    \\ 
        6     &  creat\_type\_cd          & C24                       & device\_model    \\ 
        7     &  task\_id                 & C3                        & site\_id    \\ 
        8     &  device\_size             & C26                       & app\_category    \\ 
        9     &  city                     & C18                       & C20     \\ 
       10     &  series\_group            & C11                       & C21     \\ 
       11     & device\_name              & C15                       & app\_domain     \\
       12     & spread\_app\_id           & C7                        & C18    \\
       13     & app\_second\_class        & C19                       & device\_conn\_type     \\
       14     & emui\_dev                 & C14                       & banner\_pos    \\
       15     & hispace\_app\_tags        & C17                       & C16     \\
       16     & age                       & C22                       & C1     \\
       17     & city\_rank                & C13                       & hour     \\
       18     & gender                    & C25                       & site\_domain     \\
       19     & series\_dev               & C23                       & C15     \\
       20     & inter\_type\_cd           & C9                        & device\_type     \\
       21     & net\_type                 & C6                        & C19     \\
       22     & residence                 & C20                       & site\_category     \\
       23     & -                         & C8                        & day     \\
       24     & -                          & C5                        & id     \\
       25     & -                          & C1                        & -       \\
       26     & -                          & C2                        & -        \\ 
        \multicolumn{4}{l}{(b). Numerical features.}\\
        \hline
        \#Importance  & Digix Features      & Criteo Features           & Avazu Features    \\
        1     & u\_refreshTimes     & I3                        & -         \\    
        2     & u\_feedLifeCycle    & I8                        & -         \\  
        3     & app\_score          & I9                        & -         \\  
        4     & -                    & I13                       & -         \\
        4     & -                    & I2                        & -         \\
        4     & -                    & I12                       & -         \\
        4     & -                    & I6                        & -         \\
        4     & -                    & I4                        & -         \\
        4     & -                    & I5                        & -         \\
        4     & -                    & I7                        & -         \\
        4     & -                    & I1                        & -         \\
        4     & -                    & I11                       & -         \\
        4     & -                    & I10                       & -         \\
        \hline
        \hline
       \end{tabular} 
\end{table}

CTR prediction results of DeepFM on the Digix dataset are given in Table \ref{t2}. Table \ref{t2} (a) demonstrates the general search performance outcomes. In each iteration, five categorical features, which have the lowest score, have been removed from the categorical feature set since $i$ is set to five. In addition, no numerical features are removed during the iterations since the total number of numerical features equaled $j$, which is set to three. 

We obtain the best performance result in the fourth iteration with only five features remaining, three numerical and two categorical. Specifically, the AUC of DeepFM increases to 0.78838 from 0.77044, a significant improvement of 2.33 through the general search method. Moreover, the number of features used decreases to five from 25, leading to a reduction in training time to 378.45179 seconds from 515.40333 seconds. 

Following the completion of the general search iterations, we apply a neighborhood search to the top three results (with $k$ set to three), ranked by AUC score, to identify any potential improvements in AUC. The outcomes of the neighborhood search are shown in Table \ref{t2} (b). 

For comparative purposes, we also apply a Genetic Algorithm (GA) for feature selection. Table \ref{t2} (c) demonstrates the GA performance outcomes. 

\begin{table}[H]\scriptsize   %\footnotesize \tiny \small
    \caption{CTR prediction performance outcomes of DeepFM on Digix dataset.}
    \label{t2}
    \centering
       \begin{tabular}{lllll}
        \hline
        \hline
        \multicolumn{5}{l}{(a). General search outcomes.}\\
        \hline
        \# Features & AUC     & Logloss     & Time(s) & Rank    \\
        25 (3, 22)         & 0.77004 & 0.06952     & 515.40333  & base     \\    
        20 (3, 17)         & 0.77714 & 0.06874     & 419.64680  & 2   \\  
        15 (3, 12)         & 0.76320 & 0.07092     & 357.85660  &     \\  
        10 (3, 7)          & 0.77028 & 0.07018     & 268.89307  & 3   \\  
        5  (3, 2)          & \textbf{0.78838} & 0.06602     & 378.45179  & 1  \\ 
        \multicolumn{5}{l}{(b). Neighborhood search outcomes.}\\
        \hline
        \# Features & AUC                    & Logloss     & Time(s) & Rank    \\
        6 (3, 3)    & 0.78638                & 0.06584     & 246.77620 & 1-u   \\
        5 (2, 3)    & 0.78498                & 0.06618     & 240.14998 & 1-u   \\
        4 (1, 3)    & 0.78264                & 0.06630     & 263.38909 & 1-u   \\
        4 (2, 2)    & \textbf{0.78700}       & 0.06612     & 353.38565 & 1-d   \\
        3 (1, 2)    & 0.78322                & 0.06660     & 344.76782 & 1-d   \\
        21 (3, 18)  & 0.76212                & 0.07188     & 436.99080 & 2-u   \\
        20 (2, 18)  & 0.76334                & 0.07132     & 436.15972 & 2-u   \\
        19 (1, 18)  & 0.77208                & 0.06960     & 434.37155 & 2-u   \\
        19 (2, 17)  & 0.77972                & 0.06790     & 425.57886 & 2-d   \\
        18 (1, 17)  & 0.76638                & 0.07154     & 414.56353 & 2-d   \\
        19 (3, 16)  & 0.77356                & 0.06836     & 418.00543 & 2-d   \\
        11 (3, 8)   & 0.76380                & 0.07144     & 282.41846 & 3-u   \\
        10 (2, 8)   & 0.76572                & 0.07096     & 281.62117 & 3-u   \\
         9 (1, 8)   & 0.75502                & 0.07386     & 286.26636 & 3-u   \\
         9 (2, 7)   & 0.76050                & 0.07140     & 268.83479 & 3-d   \\
         8 (1, 7)   & 0.76254                & 0.07074     & 265.57414 & 3-d   \\
         9 (3, 6)   & 0.76072                & 0.07130     & 261.92777 & 3-d   \\
        \multicolumn{4}{l}{(c). Genetic algorithm outcome.}\\
        \hline
        \# Features & AUC                    & Logloss     & Time(s)    \\
        11 (2, 9)   & 0.79464                & 0.06540     & 505.79713  \\
        \hline
        \hline
       \end{tabular} 
\end{table}

\begin{figure}[H]
    \centering
    \graphicspath{ {figs/} }
    \includegraphics[scale=0.7]{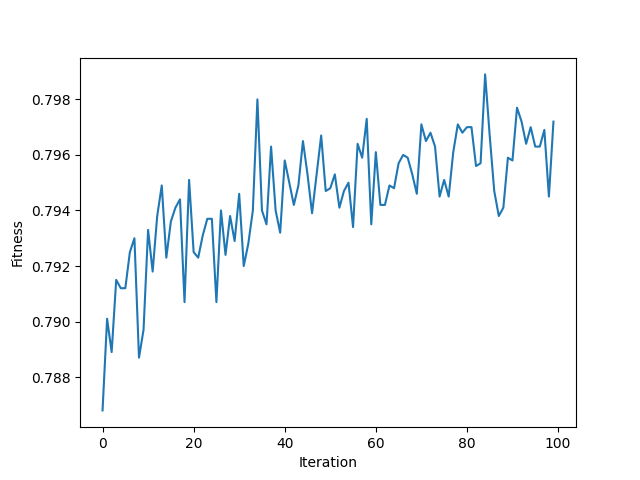}
    \caption{Evolution of fitness value over iterations on the Digix dataset.}
    \label{fig:ga1}
\end{figure}

Fig. \ref{fig:ga1} illustrates the fitness value (AUC score) of individuals in the new generation population. We defined arbitrary parameters such as population size, number of parents inside the mating pool, number of elements to mutate, and number of generations for all datasets as 8, 4, 3, and 100, respectively. The GA required 398,248.65147 seconds to find the optimal feature subset. The average AUC score of the best individual over 5 runs is provided in Table \ref{t2} (c). Although the GA achieved a slightly better AUC score than our proposed approach, its runtime is excessively long, making it impractical for use. Moreover, the process of searching for the optimal features using the GA consumes a considerable amount of computational resources. This outcome supports our motivation for developing a lightweight (robust \& fast) feature selection technique for real-world CTR prediction problems.

In Table \ref{t2}, examining the neighborhood search around the top-ranked feature set, we observe three numerical and two categorical features. During the up-neighborhood search from this top-ranked set, we initially remove three numerical features and reintroduce the last removed categorical feature. Notably, this removal results in no numerical features remaining in the subset. Consequently, we add the numerical feature with the highest feature score, which had been previously removed. In subsequent iterations of the up-neighborhood search, numerical features are added back into the subset from the removed list, one by one per iteration.

Conversely, in the down-neighborhood search, numerical features are removed from the subset one by one in each iteration. Once no numerical features remain, a categorical feature is removed, and all numerical features are reintroduced to the subset. These added numerical features are then removed sequentially, one by one, until none remain. In the Digix dataset, the down-neighborhood search is halted when the total number of features decreases to three.

Performance outcomes of DeepFM on the Criteo dataset and the Avazu dataset are presented in Table \ref{t3} and Table \ref{t4}, respectively. To further validate our proposed solution, we also employ the FiBiNET model, and the performance outcomes of the FiBiNeT model are shown in Table \ref{t5}. All results in these tables are presented as the average of the five runs to ensure robust and reliable results.

\begin{table}[H]\scriptsize   %\footnotesize \tiny \small
    \caption{ CTR prediction performance outcomes of DeepFM on Criteo dataset.}
    \label{t3}
    \centering
       \begin{tabular}{lllll}
        \hline
        \hline
        \multicolumn{5}{l}{(a). General search outcomes. }\\
        \hline
        \# Features & AUC  & Logloss & Time(s) & Rank    \\
        39 (13, 26)        & 0.73896 & 0.54980     & 637.65374 & base(3)     \\    
        36 (10, 26)        & \textbf{0.74044} & 0.54542     & 625.99047 & 1   \\  
        33 (7, 26)         & 0.74034 & 0.54716     & 582.26899 & 2    \\ 
        30 (4, 26)         & 0.73350 & 0.54678     & 565.87208 &     \\ 
        27 (1, 26)         & 0.72030 & 0.56080     & 560.91164 &     \\ 
        34 (13, 21)        & 0.73536 & 0.56184     & 536.85893 &     \\ 
        31 (10, 21)        & 0.73606 & 0.55520     & 531.78017 &     \\ 
        28 (7, 21)         & 0.73202 & 0.55738     & 516.96770 &     \\ 
        25 (4, 21)         & 0.72836 & 0.56224     & 501.91147 &     \\ 
        22 (1, 21)         & 0.72420 & 0.55846     & 493.26440 &     \\ 
        29 (13, 16)        & 0.72966 & 0.56838     & 469.92229 &     \\ 
        26 (10, 16)        & 0.72840 & 0.57514     & 475.92175 &     \\ 
        23 (7, 16)         & 0.73374 & 0.54816     & 469.19922 &     \\ 
        20 (4, 16)         & 0.71592 & 0.58208     & 451.60715 &     \\ 
        17 (1, 16)         & 0.71836 & 0.56464     & 439.61538 &     \\ 
        24 (13, 11)        & 0.72928 & 0.57056     & 409.26729 &     \\ 
        21 (10, 11)        & 0.71650 & 0.58086     & 399.49282 &     \\ 
        18 (7, 11)         & 0.72784 & 0.55010     & 391.41615 &     \\
        15 (4, 11)         & 0.70022 & 0.59502     & 385.86003 &     \\
        12 (1, 11)         & 0.70752 & 0.58074     & 373.78389 &     \\
        19 (13, 6)         & 0.71368 & 0.56098     & 306.53366 &     \\
        16 (10, 6)         & 0.70908 & 0.57088     & 297.79642 &     \\
        13 (7, 6)          & 0.71258 & 0.56984     & 291.11090 &     \\
        10 (4, 6)          & 0.70132 & 0.56848     & 282.44822 &     \\
         7 (1, 6)          & 0.68010 & 0.57850     & 267.99245 &     \\
        14 (13, 1)         & 0.69880 & 0.57528     & 219.52942 &     \\
        11 (10, 1)         & 0.69426 & 0.57180     & 194.72427 &     \\
         8 (7, 1)          & 0.65742 & 0.64306     & 184.44831 &     \\
         5 (4, 1)          & 0.67132 & 0.58724     & 160.62160 &     \\
         2 (1, 1)          & 0.62956 & 0.6331      & 150.92957 &     \\            
        
        \multicolumn{5}{l}{(b). Neighborhood search outcomes.}\\
        \hline
        \# Features             & AUC     & Logloss     & Time(s) & Iteration    \\
        %39                     & 0.74030 & 0.54874     & 619.67152 & base(3)     \\     
        38 (12, 26)         & \textbf{0.74432} & 0.54624     & 658.56682 & 1-u   \\   
        37 (11, 26)         & 0.74042 & 0.54764     & 661.28652 & 1-u   \\ 
        %36          & 0.74118 & 0.54658     & 605.43219 & 1   \\
        35 (9, 26)          & 0.73542 & 0.55806     & 599.21552 & 1-d   \\
        34 (8, 26)          & 0.73014 & 0.57284     & 609.86739 & 1-d   \\
        %33          & 0.73516 & 0.55902     & 586.83153 & 2   \\
        32 (6, 26)          & 0.73116 & 0.55718     & 572.16763 & 2-d    \\
        31 (5, 26)          & 0.72628 & 0.56040     & 559.99307 & 2-d    \\
        \multicolumn{4}{l}{(c). Genetic algorithm outcome.}\\
        \hline
        \# Features & AUC        & Logloss     & Time(s)    \\
        25 (10, 15) & 0.77518    & 0.47438     & 502.03056  \\
        \hline
        \hline
        \hline
       \end{tabular} 
\end{table}

Table \ref{t3} (a) shows that the best three AUC scores are obtained with a total of 36, 33, and 39 features, respectively. It is observed that most of the general search iterations do not improve the AUC score compared to the baseline result. Notably, the baseline result ranks third in terms of AUC score. Additionaly, Table \ref{t3} (b) demonstrates that the best AUC score is achieved in the second iteration of the up-search for the first-ranked result. The overall AUC score increases significantly to 0.74432 from 0.73896, representing a 0.7253\% improvement. Moreover, the overall Logloss score decreases to 0.54624 from 0.54980. 

In terms of training time, there is an approximately 20-second difference between the best result and the baseline result. However, if the training time increases despite a reduction in the number of features, this may indicate model overfitting due to the limited amount of data used in the experiments, which is constrained by hardware limitations. Therefore, training time is a critical metric in our experiments, as we aim to reduce the time required to train the model with a decreased number of features.

\begin{figure}[H]
    \centering
    \graphicspath{ {figs/} }
    \includegraphics[scale=0.7]{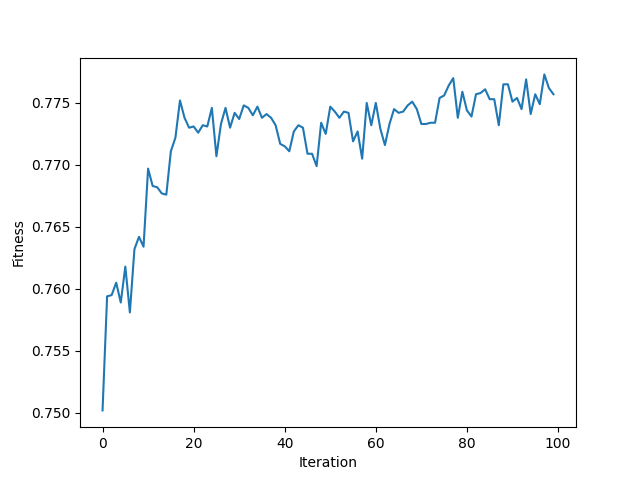}
    \caption{Evolution of fitness value over iterations on the Criteo dataset.}
    \label{fig:ga2}
\end{figure}

Additionally, Fig. \ref{fig:ga2} depicts the GA spending 361,927.67209 seconds to find the optimal feature subset on the Criteo dataset.

\begin{table}[H]\scriptsize   %\footnotesize \tiny \small
    \caption{CTR prediction performance outcomes of DeepFM on Avazu dataset.}
    \label{t4}
    \centering
       \begin{tabular}{lllll}
        \hline
        \hline
        \multicolumn{5}{l}{(a). General search outcomes.}\\
        \hline
        \# Features & AUC     & Logloss     & Time(s) & Rank    \\
        24          & 0.74512 & 0.44246     & 521.05655 & base     \\    
        19          & \textbf{0.75996} & 0.41330     & 419.66187 & 1   \\  
        14          & 0.75776 & 0.42340     & 360.43661 & 2    \\ 
         9          & 0.75468 & 0.42848     & 288.22345 & 3    \\ 
         4          & 0.73574 & 0.42944     & 176.40335 &     \\        
        
        \multicolumn{5}{l}{(b). Neighborhood search outcomes.}\\
        \hline
        \# Features & AUC     & Logloss     & Time(s) & Iteration    \\
        %24          & 0.74962 & 0.45674     & 544.22794  & base     \\     
        22          & 0.75392 & 0.43464     & 550.40292 & 1-u   \\   
        21          & 0.75642 & 0.42258     & 534.72531 & 1-u   \\ 
        20          & \textbf{0.75866}      & 0.42172     & 471.62775 & 1-u   \\
        %19          & 0.76084 & 0.42094     & 444.88186 & 1   \\
        18          & 0.75140 & 0.42934     & 418.67870 & 1-d   \\
        17          & 0.75272 & 0.42516     & 408.96943 & 1-d   \\
        16          & 0.75304 & 0.43258     & 410.49349 & 1-d   \\
        15          & 0.75196 & 0.43812     & 393.04971 & 2-u   \\
        %14          & 0.75538 & 0.43104     & 375.51440 & 2   \\
        13          & 0.75430 & 0.42402     & 366.12281 & 2-d   \\
        12          & 0.75472 & 0.42750     & 350.57764 & 2-d   \\
        11          & 0.75372 & 0.43174     & 332.01572 & 2-d   \\
        10          & 0.75536 & 0.42688     & 311.55026 & 3-u    \\
         %9          & 0.75624 & 0.42336     & 301.14609 & 3    \\
         8          & 0.74584 & 0.44174     & 289.80166 & 3-d    \\
         7          & 0.75280 & 0.42620     & 272.75906 & 3-d    \\
         6          & 0.74126 & 0.43108     & 248.76237 & 3-d    \\
        \multicolumn{4}{l}{(c). Genetic algorithm outcome.}\\
        \hline
        \# Features & AUC        & Logloss     & Time(s)    \\
        13          & 0.74968    & 0.43346     & 427.21833  \\
        \hline
        \hline
        \hline
       \end{tabular} 
\end{table}

Table \ref{t4} presents the Avazu dataset, which comprises 24 categorical features without any numerical features. In the general search (a), five features are iteratively removed until fewer than five features remain in each iteration. The resulting subsets of 19, 14, and 9 features outperform the base (24 features) by 1.99 \%, 1.70 \%, and 1.28 \%, respectively, in terms of AUC score. Moreover, they exhibit superior performance in Logloss score and Time metric, surpassing the base by 6.59 \%, 3.31 \%, and 3.26 \%, and 19.45 \%, 30.83 \%, and 44.68 \%, respectively. 

Results are further ranked by AUC score, as indicated in the rank column, and subjected to a neighborhood search applied to the top three results (b), as shown in Table \ref{t4}. Notably, all results, except the last one featuring 6 features, demonstrate improved AUC scores compared to the base (24 features). Despite the slightly lower AUC score of the last result compared to the base, its Logloss score and Time metric are superior. Additionally, all neighborhood results outperform the base in terms of Logloss and Time.

However, none of the neighborhood results surpass the first-ranked result in terms of AUC score, with the subset of 19 features yielding the best performance overall.

\begin{figure}[H]
    \centering
    \graphicspath{ {figs/} }
    \includegraphics[scale=0.7]{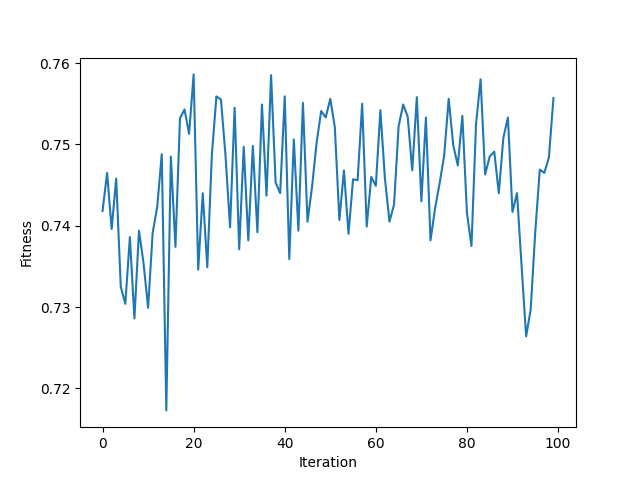}
    \caption{Evolution of fitness value over iterations on the Avazu dataset.}
    \label{fig:ga3}
\end{figure}

Additionally, Fig. \ref{fig:ga3} depicts the GA spending 96,985.12194 seconds to find the optimal feature subset on the Avazu dataset.

Table \ref{t5} reveals the superiority of the proposed solution. In the general search (a), all results outperform the base in terms of all metrics. Specifically, a subset of 10 features (3 numerical and 7 categorical) demonstrates better performance than the base (25 features) in terms of AUC, Logloss, and Time metrics by 3.80 \%, 9.28 \%, and 79.72 \%, respectively. Similarly, a subset of 15 features shows improved performance over the base in terms of AUC, Logloss, and Time metrics by 3.23 \%, 7.59 \%, and 60.42 \%, respectively. Additionally, a subset of 5 features exhibits better performance than the base in terms of AUC, logloss, and time metrics by 3.22 \%, 10.66 \%, and 82.50 \%, respectively.

Moreover, the best result is achieved in the neighborhood search (b), where only 11 features (3 numerical and 8 categorical) are utilized, leading to the highest AUC score. In this case, significant improvements are observed compared to the base, with enhancements of 4.17 \%, 9.85 \%, and 76.72 \% in terms of AUC, Logloss, and Time, respectively.   

\begin{table}[H]\scriptsize   %\footnotesize \tiny \small
    \caption{ CTR prediction performance outcomes of FiBiNet on Digix dataset.}
    \label{t5}
    \centering
       \begin{tabular}{lllll}
        \hline
        \hline
        \multicolumn{5}{l}{(a). General search outcomes. }\\
        \hline
        \# Features & AUC     & Logloss     & Time(s) & Rank    \\
        25 (3, 22)         & 0.76340 & 0.07374     & 3137.90468 & base     \\    
        20 (3, 17)         & 0.77538 & 0.07002     & 2124.28965 &    \\  
        15 (3, 12)         & 0.78806 & 0.06814     & 1242.11011 & 2    \\  
        10 (3, 7)          & \textbf{0.79238} & 0.06690     & 636.25120  & 1    \\  
        5  (3, 2)          & 0.78798 & 0.06588     & 549.06432  & 3    \\ 
        \multicolumn{5}{l}{(b). Neighborhood search outcomes.}\\
        \hline
        \# Features & AUC               & Logloss     & Time(s) & Iteration    \\
        11 (3, 8)    & \textbf{0.79520}  & 0.06648     & 730.56722 & 1-u   \\
        10 (3, 8)    & 0.79326 & 0.06792     & 715.76065 & 1-u   \\
        9 (1, 8)     & 0.76858 & 0.07152     & 704.34727 & 1-u   \\     
        9 (2, 7)     & 0.78944 & 0.06784     & 607.28292 & 1-d   \\
        8 (1, 7)     & 0.79044 & 0.06796     & 704.97161 & 1-d   \\
        9 (3, 6)     & 0.79246 & 0.06672     & 704.09572 & 1-d   \\
        16 (3, 13)   & 0.78622 & 0.06788     & 1410.03851 & 2-u  \\
        15 (2, 13)   & 0.78432 & 0.06800     & 1425.68767 & 2-u  \\
        14 (1, 13)   & 0.77200 & 0.07176     & 1391.41916 & 2-u  \\
        14 (2, 12)   & 0.77620 & 0.06922     & 1217.43596 & 2-d  \\
        13 (1, 12)   & 0.77504 & 0.06968     & 1188.06378 & 2-d  \\
        14 (3, 11)   & 0.76686 & 0.07220     & 1078.11548 & 2-d  \\
        6 (3, 3)     & 0.78570 & 0.06590     & 520.71559 & 3-u  \\
        5 (2, 3)     & 0.78466 & 0.06616     & 460.72673 & 3-u  \\
        5 (2, 3)     & 0.77698 & 0.06670     & 460.96362 & 3-u  \\
        4 (1, 3)     & 0.78806 & 0.06614     & 496.15163 & 3-d  \\
        4 (1, 2)     & 0.78140 & 0.06674     & 498.96149 & 3-d  \\
        \hline
        \hline
       \end{tabular} 
\end{table}

\section{Discussion}
\label{discussion}

The proposed NeSHFS method has shown promising results in the context of Click-Through Rate (CTR) prediction across multiple datasets. Our evaluations, conducted using the DeepCTR framework and the DeepFM and FiBiNeT models, highlight the method's effectiveness in both feature selection and model performance enhancement.

The NeSHFS method demonstrates a significant improvement in prediction accuracy, as evidenced by the AUC scores. For instance, on the Huawei Digix 2022 dataset, the AUC score increased from 0.77044 to 0.78838, a notable 2.33\% improvement, while reducing the number of features from 25 to 5. This reduction also led to a decrease in training time from 515.40333 seconds to 378.45179 seconds, showcasing the method's efficiency in handling large datasets with fewer computational resources. 

While the Genetic Algorithm (GA) yielded slightly better AUC scores, its runtime was substantially longer, making it impractical for real-world applications. The GA required over 398,000 seconds for feature selection on the Digix dataset, compared to NeSHFS, which achieved faster results with less computational overhead. This demonstrates NeSHFS's advantage as a lightweight and robust feature selection technique.

Our experiments revealed distinct patterns in feature importance across datasets. For example, in the Digix dataset, slot\_id, C12, and device\_id were the most critical categorical features, while u\_refreshTimes and I3 were the most influential numerical features. This distinction underscores the necessity of tailored feature selection processes for different types of data, which NeSHFS effectively addresses through its dual-ranking system.

The general search process effectively identified high-performing feature subsets by iteratively removing the least important features. The subsequent neighborhood search further refined these subsets, leading to even higher AUC scores. This two-step approach proved beneficial in maximizing model performance and ensuring that the feature sets were optimized for each specific dataset.

NeSHFS's performance across different datasets (Huawei Digix 2022, Criteo, and Avazu) demonstrates its scalability and generalizability. The method's adaptability to various data types and sizes makes it a valuable tool for CTR prediction tasks across diverse domains.

While NeSHFS performed well across the evaluated datasets, its performance may vary with different types of data and CTR prediction models. Future work should explore its applicability to other datasets and contexts to further validate its robustness and versatility.

The performance of NeSHFS depends on several parameters, such as the number of features to remove or add during the search processes. While our experiments used arbitrarily chosen values, a more systematic approach to parameter tuning could potentially enhance the method's performance.

NeSHFS was validated using DeepFM and FiBiNeT models. Future research could explore its integration with other advanced deep learning models to assess its effectiveness in broader machine learning applications.

In summary, the NeSHFS method presents a compelling approach to feature selection for CTR prediction, combining efficiency, effectiveness, and adaptability. Its ability to deliver significant performance improvements with reduced computational resources positions it as a valuable tool for real-world CTR prediction tasks.

\section{Conclusions}
\label{conclusions}
In this paper, we introduced NeSHFS, a heuristic-based feature selection method. Our approach involves ranking numerical features by their ANOVA scores and categorical features by their chi-square scores. NeSHFS iteratively selects and removes an appropriate number of features based on their decreasing ranking scores, both from numerical and categorical feature sets. These selected feature subsets are then employed in CTR prediction using a chosen model and dataset. The feature sets yielding the highest prediction results undergo further refinement through a local search mechanism. This involves iteratively adding and removing features from the feature set while applying the same CTR prediction process. Our solution was validated on three distinct datasets, with experimental results demonstrating significant improvements in AUC scores and significant reductions in execution time.

\section*{Acknowledgements}
This work is supported by METU with grant no ADEP-312-2024-11484.

\bibliography{sample}

\end{document}